\documentclass{iopart}

\usepackage{iopams}
\usepackage{hyperref,graphicx,subfig}

\newcommand{\imag}{\Im\hspace{-0.1pt}\mathfrak{m}\hspace{0.1pt}}

\begin{document} 

\title{Kadanoff-Baym approach to double-excitations in finite systems}
\author{N S\"{a}kkinen, M Manninen, and R van Leeuwen}
\address{Department of Physics, Nanoscience Center, FIN 40014, University of Jyv\"{a}skyl\"{a}, Jyv\"{a}skyl\"{a}, Finland}
\ead{niko.sakkinen@jyu.fi}

\begin{abstract}
We benchmark many-body perturbation theory by studying neutral, as well as non-neutral, excitations of finite lattice systems.  The neutral excitation spectra are obtained by time-propagating the Kadanoff-Baym equations in the Hartree-Fock and second Born approximations. Our method is equivalent to solving the Bethe-Salpeter equation with a high-level kernel while respecting self-consistently, which guarantees the fulfillment of a frequency sum rule. As a result, we find that a time-local method, such as Hartree-Fock, can give incomplete spectra, while already the second Born, which is the simplest time-nonlocal approximation, reproduces well most of the additional excitations, which are characterized as double-excitations.
\end{abstract}
\pacs{31.10.+z,71.10.-w, 31.15.xm}


   \section{Introduction}
   We recently developed a non-equilibrium many-body method, based on time-propagation of the Kadanoff-Baym equations (KBE), to study transient dynamics in quantum transport, and found that electronic correlations beyond mean-field strongly affect the relevant non-equilibrium properties~\cite{Myohanen_EPL84,Myohanen_PRB80}. The method is approximate, because only some selected classes of perturbative terms are accounted for, and therefore an estimate of the quality of the results would be desirable. However, as benchmark results are scarcely available for such open systems, we need to devise simpler experiments to test our method. Some finite, exactly solvable lattice systems, which typically model molecular devices and whose properties thus determine the transport characteristics, form an appropriate test platform. A many-body approximation which is unable to reproduce the properties of such an isolated system correctly, has no prerequisites to fare well for a molecular device between conducting leads.

   The Kadanoff-Baym equations are equations of motion for the one-body Green's function $G$, where many-body effects are incorporated into the so-called self-energy $\Sigma$. The latter is a functional of the Green's function, and the main task of many-body perturbation theory (MBPT) is to find good approximations for this quantity. We are, in particular, interested in the so-called conserving approximations for the self-energy, which are vital in quantum transport~\cite{Thygesen1,Strange1} as they guarantee the fulfillment of the conservation laws for the particle number, total momentum and energy~\cite{Baym_PR2,Baym_PR4}. Some of these approximations have already been tested in finite lattice systems~\cite{PuigVonFriesen1,PuigVonFriesen2}. These reports have highlighted strongly perturbed systems, in which serious issues, such as artificial damping of the dynamics, were observed and deemed correlation-induced as they were not present in a mean-field approximation. The purpose of this paper is to extend these investigations to the linear response regime. 

   Within this regime, the key quantity is the response function $\delta G/\delta v$ which describes the change in the Green's function due to a weak external field $v$, and gives a direct access to the neutral excitation spectra. This quantity satisfies an integral equation known as the Bethe-Salpeter equation (BSE)~\cite{Strinati_RDNC11}, whose integral kernel is given by the functional derivative $\delta \Sigma/ \delta G$. Using this equation to calculate the excitation spectra has proven to be computationally challenging and, in practice, often requires a number of additional approximations, such as neglect of self-consistency, kernel diagrams, and/or frequency-dependence. Instead, we obtain the response function by time-propagation of the Green's function which does not require any of the above mentioned approximations~\cite{Kwong_PRL84,Dahlen2}. Moreover, such obtained excitation spectra automatically satisfy a frequency sum rule known as the f-sum rule, which acts as an important consistency relation for the response function.

   The spectral properties of neutral, as well as non-neutral, excitations depend strongly on the temporal structure of the self-energy which can be either time-local or -nonlocal. Whereas a time-local scheme leads to merely renormalized one-particle excitations, a time-nonlocal one can reproduce more complicated spectral structure. Such excitations involving double- or many-particle character are, in fact, dominant for some potential materials for realistic devices, such as organic polymers (polyenes)~\cite{Starcke1}, making time-nonlocal approximations an interesting research topic. This represents a major challenge for the Bethe-Salpeter approach as time-nonlocality translates to an integral kernel which depends on three frequencies. Although some state-of-the-art calculations have been recently conducted in finite systems with a time-nonlocal kernel~\cite{Romaniello_JCP130, Sangalli_JCP134}, these results have become at the expense of self-consistency which is a requirement for fulfillment of conservation laws, and therefore indispensable in a time-dependent transport process. Moreover, a non-self-consistent approximation can lead to a gross violation of the f-sum rule~\cite{Kwong_PRL84,Pal1}. On the other hand, our calculations, conducted within the time-local Hartree-Fock (HF) and time-nonlocal second Born (2nd Born, 2B) approximations~\cite{Dahlen2,Dahlen1}, do not sacrifice self-consistency, remain computationally tractable, and yield excitation spectra satisfying the f-sum rule.

   The paper is organized as follows. We start with an introductory section~\ref{sec:mbpt} in which we summarize the non-equilibrium many-body formalism, describe our means to approach the neutral excitation spectra, and introduce, as well as analyze, our many-body approximations. In section~\ref{sec:model}, we introduce the test environment which comprises two simple lattice systems. Our numerical results, in particular, on addition, removal, and excitation spectra, as well as some related discussion are contained in section~\ref{sec:results}. We summarize our work and present some conclusions in section~\ref{sec:conclusions}. Finally,~\ref{sec:proof_of_fsum_rule} contains an extension of the proof the f-sum rule to lattice systems.    


   \section{Response functions in non-equilibrium many-body theory}
   \label{sec:mbpt}


      \subsection{Non-equilibrium Kadanoff-Baym approach}
\begin{figure}[b]
         \centering
         \includegraphics[scale=1.0]{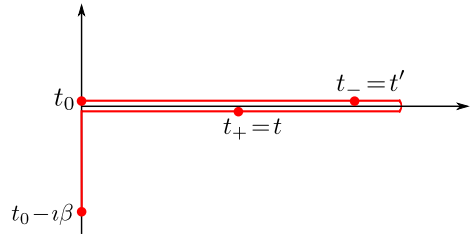}
         \caption{The extended Keldysh contour starting from $t_{0}$ and ending to $t_{0} - 
mi\beta$ consists of an equilibrium, imaginary track and a two non-equilibrium, real-time branches. The zero-temperature limit $\beta\rightarrow\infty$, where $\beta$ is the inverse temperature, is implied when finite systems are considered.}
         \label{fig:keldysh_contour}
      \end{figure}

      The Green's functions are reduced quantities which act as probes to physical observables, while containing only a minimal amount of excess information. The non-equilibrium one-body Green's function is defined as 
      \begin{eqnarray}
      \label{eq:1GF}
         G_{ij}(z,z') \equiv -\rmi\big<\mathcal{T}_{\gamma}\hat{c}_{H,i}(z)\hat{c}_{H,j}^{\dagger}(z')\big>\, ,
      \end{eqnarray}
      where $\langle\dots\rangle$ denotes an ensemble average with respect to the grand canonical density operator, and $\hat{c}_{H,i}^{(\dagger)}(z)$ a Heisenberg picture operator which annihilates (creates) a particle from (to) a single-particle quantum state $i$. The contour-ordering operator $\mathcal{T}_{\gamma}$ arranges the contour times $z$ and $z'$ along the extended Keldysh contour $\gamma$~\cite{Keldysh} shown in figure~\ref{fig:keldysh_contour}, for further details see~\cite{Danielewicz1,VanLeeuwen1}. 

      The contour-ordered Green's function, although a powerful formal device, does not represent directly any observable. It can, however, be reduced to a set of more intuitive and physical objects by exposing all possible time-orderings. The one-body Green's function, being a two-time object, can then be written as
      \begin{eqnarray}
         \boldsymbol{G}(z,z') = \theta(z,z')\boldsymbol{G}^{>}(z,z') + \theta(z',z)\boldsymbol{G}^{<}(z,z')\, ,
      \end{eqnarray}
      where the boldfaced symbols denote matrices in the basis spanned by the single-particle quantum states. The greater and lesser Keldysh components, which are defined as
      \numparts
         \begin{eqnarray}
            G_{ij}^{>}(z,z') &\equiv -\rmi\big<\hat{c}_{H,i}(z)\hat{c}_{H,j}^{\dagger}(z')\big>\, , \\
            G_{ij}^{<}(z,z') &\equiv \rmi\big<\hat{c}_{H,j}^{\dagger}(z')\hat{c}_{H,i}(z)\big>\, ,
         \end{eqnarray}
      \endnumparts 
      describe the motion of a particle and a hole, respectively, in the many-particle system. The core of the non-equilibrium theory lies in its equations of motion, the first of which can be written as
      \begin{eqnarray}
         \label{eq:kadanoff_baym}
         \big(\rmi\partial_{z} - \boldsymbol{h}(z)\big) \boldsymbol{G}(z,z')
         = \boldsymbol{\delta}(z,z') + \int\limits_{\gamma}\!d\bar{z}\; \boldsymbol{\Sigma}[\boldsymbol{G}](z,\bar{z}) \boldsymbol{G}(\bar{z},z')\, ,
      \end{eqnarray}
      while the second, adjoint equation can be obtained by differentiation with respect to the second time-argument. The symbol $\boldsymbol{h}(z)$ denotes the one-body part of the Hamiltonian which can, in general, be time-dependent. Using the Langreth rules~\cite{Langreth1,VanLeeuwen1}, the equations of motion can be split into a set of equations for the different Keldysh components referred as the Kadanoff-Baym equations~\cite{KadanoffBaym,Danielewicz1}. The self-energy $\Sigma$ is an effective, nonlocal potential, which accounts for all the many-body effects, and allows a systematic, beyond order-by-order perturbation expansion. A Green's function obtained from an approximate self-energy
      \begin{eqnarray}
         \Sigma_{ij}(z,z') = \frac{\delta \Phi[G]}{\delta G_{ji}(z',z)}\, ,
      \end{eqnarray}
      where $\Phi[G]$ is the Baym-functional~\cite{Baym_PR4}, obeys the conservation laws for the particle number, total momentum and energy provided that it satisfies the equation of motion~\eref{eq:kadanoff_baym}, or in other words is solved self-consistently~\cite{Baym_PR2,Baym_PR4}. 

      The nonlocal structure of the one-body Green's function ensures that all one-body observables which are accessible in the equilibrium/zero-temperature theory can be now calculated in non-equilibrium settings from the knowledge of the greater and lesser components. In particular, all time-local observables are related to the one-body reduced density matrix (1RDM) which is given by
      \begin{eqnarray}
      \label{eq:density_matrix}
         \boldsymbol{\gamma}(t) \equiv -\rmi \boldsymbol{G}^{<}(t,t)\, ,
      \end{eqnarray}
      where $t$ is a real-time argument located on the horizontal track of the Keldysh contour. On the other hand, time-nonlocality allows access to information on particle number changing processes contained in the spectral function, which is defined as
      \begin{eqnarray}
      \label{eq:spectral_function}
         \boldsymbol{A}(t,t') \equiv \rmi \big[ \boldsymbol{G}^{>}(t,t') - \boldsymbol{G}^{<}(t,t')\big]\, ,
      \end{eqnarray}
      and whose pole structure in the frequency-domain gives the particle addition and removal energies. Another virtue of time-nonlocality, is the possibility to access some time-local two-body observables, including the total energy which can be written in terms of the Galitski-Migdal (GM) functional as 
      \begin{eqnarray}
      \label{eq:galitskimigdal}
         E(t) = -\frac{\rmi}{2}\tr \Big[\big( \boldsymbol{\partial}_{t} + \boldsymbol{h}\big) \boldsymbol{G}^{<}(t,t')\Big]_{t'=t} \, .
      \end{eqnarray}
      where $\boldsymbol{\partial}_{t}$ is proportional to the identity matrix, and $\tr\boldsymbol{A} \equiv \sum_{i} A_{ii}$ is the regular algebraic trace. However, the main advantage of the non-equilibrium theory addressed here, is the additional possibility to calculate some time-nonlocal two-body observables from the sole knowledge of the one-body Green's function. We focus on the real-time, retarded response function, which is defined as the functional derivative
      \begin{eqnarray}
      \label{eq:retarded_response_function}
         \chi_{ij,kl}^{R}(t,t') \equiv \frac{\delta \gamma_{ij}(t)}{\delta v_{lk}(t')}\bigg|_{v=0}
      \end{eqnarray}
      with respect to an external perturbation $v_{ij}(t)$, and has the general structure of a retarded function given by
      \begin{eqnarray}
      \label{eq:retarded_response_function_structure}
         \chi_{ij,kl}^{R}(t,t') = \theta(t-t') \big(\chi_{ij,kl}^{>}(t,t') - \chi_{ij,kl}^{<}(t,t')\big)\, ,
      \end{eqnarray}
      where the greater and lesser components are defined, respectively, in terms of the one-body reduced density matrix operator $\hat{\gamma}_{H,ij}(t) \equiv \hat{c}_{H,j}^{\dagger}(t)\hat{c}_{H,i}(t)$, as
      \numparts
      \label{eq:components_of_response_function}
         \begin{eqnarray}
            \chi_{ij,kl}^{>}(t,t') &\equiv -\rmi\big<\hat{\gamma}_{H,ij}(t)\hat{\gamma}_{H,kl}(t')\big>\, , \\
            \chi_{ij,kl}^{<}(z,z') &\equiv -\rmi\big<\hat{\gamma}_{H,kl}(t')\hat{\gamma}_{H,ij}(t)\big>\, .
         \end{eqnarray} 
      \endnumparts
      The retarded response function, therefore, describes the motion of a particle-hole pair, or in other words an excitation, in the many-particle system. This quantity is connected to the one-body Green's function through the linear response relation which can be written as
      \begin{eqnarray}
      \label{eq:linear_response_relation}
         \delta \gamma_{ij}(t) = \sum_{kl}\int\limits_{t_{0}}^{\infty}\!d\bar{t}\; \chi_{ij,lk}^{R}(t,\bar{t}) v_{kl}(\bar{t})\, ,
      \end{eqnarray}
      where $\delta \gamma_{ij}(t)$ is the first-order variation of the one-body reduced density matrix. The lowest order response of the system to an external perturbation is, therefore, uniquely determined by the real-time, retarded response function, whose pole structure in the frequency-domain gives the neutral excitation energies. In the non-equilibrium approach, both the 1RDM and perturbation are known, and therefore the response function is readily obtained by a proper choice of perturbation, and subsequent inversion of equation~\eref{eq:linear_response_relation}~\cite{Kwong_PRL84,Dahlen2}.

 
      \subsection{Generalized response function}
\begin{figure}[t]
         \centering
         \includegraphics[scale=1.0]{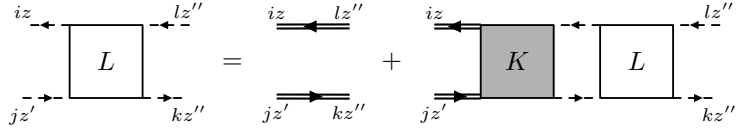}
         \caption{The Bethe-Salpeter equation for the generalized response function whose particular time-ordering $L_{ij,kl}(zz^{+},z')$ known as the particle-hole propagator yields the retarded response function. The dashed lines represent possible connections, while double lines denote dressed Green's functions.}
         \label{fig:bethe_salpeter_equation}   
      \end{figure}

      The quality of the one-body Green's function and, because of equation~\eref{eq:linear_response_relation}, also of the retarded response function is clearly determined by the self-energy approximation. Here we write down an explicit relation between the self-energy and retarded response function. We start by introducing the generalized, contour-ordered response function, which is defined as
      \begin{eqnarray}
      \label{eq:generalized_response_function}
         L_{ij,kl}(zz',z'') \equiv \frac{\delta G_{ij}(z,z')}{\delta v_{lk}(z'')}\bigg|_{v=0}
      \end{eqnarray}
      and in terms of which, the greater and lesser components of the real-time response function, defined in equation~\eref{eq:components_of_response_function}, are given by
      \begin{eqnarray}
      \label{eq:generalized_retarded_response_functions}
         \chi_{ij,kl}^{\gtrless}(t,t') = -\rmi L_{ij,kl}(zz^{+},z')\bigg|_{z=t_{\pm}, z'=t_{\mp}'}
      \end{eqnarray}
      where $z^{+} \equiv z + \eta$ with the limit $\eta \rightarrow 0^{+}$ implied after contour-ordering. The contour-times $z=t_{\pm}$ are located on the lower/upper horizontal branch of the Keldysh contour shown in figure~\ref{fig:keldysh_contour}, and correspond to a real-time $t$. The generalized response function can be straightforwardly shown~\cite{Strinati_RDNC11} to obey the so-called Bethe-Salpeter equation which is given by
      \begin{eqnarray}
         \label{eq:bethe_salpeter}
         \fl L_{ij,kl}(zz',z'') = G_{il}(z,z'')G_{kj}(z'',z') \nonumber\\
         \hspace{-2.0cm} + \sum_{pqrs}\int\limits_{\gamma}\!d(z_{1}z_{2}z_{3}z_{4})\; G_{ip}(z,z_{1})G_{qj}(z_{2},z') K_{pq,sr}(z_{1}z_{2},z_{4}z_{3})L_{rs,kl}(z_{3}z_{4},z'') \, ,
      \end{eqnarray}
      where we defined the four-point Bethe-Salpeter kernel as the functional derivative
      \begin{eqnarray}
         K_{ij,kl}(zz',\bar{z}\bar{z}') \equiv \frac{\delta \Sigma_{ij}(z,z')}{\delta G_{lk}(\bar{z}',\bar{z})}\, .
      \end{eqnarray}
      Figure~\ref{fig:bethe_salpeter_equation} contains a diagrammatic representation of this equation, and illustrates the role of the kernel as a source of an infinite perturbation series. We can conclude that any retarded response function obtained by inversion of equation~\eref{eq:linear_response_relation}, and relating to a one-body Green's function for an approximate self-energy $\Sigma$, corresponds to a solution of the Bethe-Salpeter equation with the kernel $\delta\Sigma/\delta G$. The correspondence further implies that a non-equilibrium approach already with a relative simple self-energy, accounts to a quite sophisticated approximation for the response functions.

      \subsection{Approximate Bethe-Salpeter kernels}
      \label{sec:approximations}
      The time-propagation of a Green's function with an approximate self-energy $\Sigma$ automatically leads to a retarded response function related to an approximate kernel $\delta \Sigma/\delta G$ which, for example, guarantees the obedience of the f-sum rule, as proven in~\ref{sec:proof_of_fsum_rule}. Such consistency ensures that each self-energy diagram containing $n$ Green's function lines, gives rise to $n$ inequivalent kernel diagrams which contribute to the response function. Our two approximate kernels are given by functional derivatives of the Hartree-Fock and 2nd Born self-energies, and represent a time-local mean-field and a nonlocal correlated approximation, respectively.

      \begin{figure}[b]
         \centering
         \subfloat[Hartree-Fock: self-energy $\Sigma_{HF}$ and Bethe-Salpeter kernel $K_{HF} = \delta \Sigma_{HF}/G_{HF}$.]
         {
            \centering
            \includegraphics[scale=1.0]{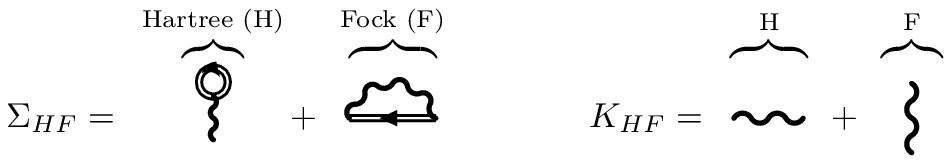}
            \label{fig:hartree_fock}
         }
         \vskip 0.0cm
         \subfloat[2nd Born: self-energy $\Sigma_{2B}$ and Bethe-Salpeter kernel $K_{2B} = \delta \Sigma_{2B}/G_{2B}$.]
         {
            \centering
            \includegraphics[scale=1.0]{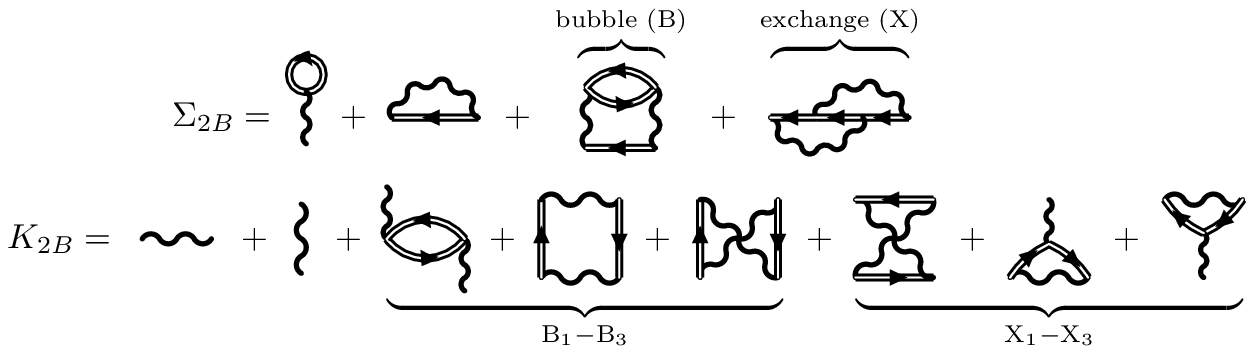}
            \label{fig:second_born}
         }
         \caption{Self-energy diagrams give rise to kernel diagrams, for example bubble (B) gives diagrams B$_{1-3}$. Wiggly lines denote the Coulomb interaction, and double lines dressed Green's functions.}
         \label{fig:approximations}
      \end{figure}

      We can view our generalized response functions diagrammatically by inserting the approximate kernels into the Bethe-Salpeter equation shown in figure~\ref{fig:bethe_salpeter_equation}. Guided by this recipe, we easily deduce that the Hartree-Fock kernel shown in figure~\ref{fig:hartree_fock} yields a response function given by a simple bubbles-and-ladders series. Whereas the second Born kernel, which is given in figure~\ref{fig:second_born}, results into a considerably more complicated response function. Note that, our approximations, whether related to a self-energy or a kernel, are known by the name of the self-energy without any extensions. For example, our Hartree-Fock approximation for the kernel shown in figure~\ref{fig:hartree_fock} is often explicitly referred to as the time-dependent Hartree-Fock approximations. 

      We can also assign a meaning to each diagram in these perturbation expansions by considering a particular time-ordering which, in our case, is the ordering given by the particle-hole propagator $L_{ij,kl}(zz^{+},z')$. We interpret our kernel diagrams as follows. The first order kernel diagrams denoted as H and F represent either a direct process (F) in which a particle-hole pair enters, interacts and exits the event, or an exchange process (H) in which a particle-hole pair enters, annihilates, and creates another pair which then exists the event. The second order kernel diagrams $B_{1-3}$ and $X_{1-3}$ can be deciphered similarly. Diagram $B_{1}$ describes a partially screened interaction between a particle and a hole, while diagrams $B_{2}-B_{3}$ reflect the exchange symmetry, or describe processes in which the original particle-hole pair is annihilated, and another emerges the event. Diagrams $X_{1-3}$ relate only to partial exchange processes in which either an entering particle ($X_{1,2}$) or hole ($B_{1,3}$) gets annihilated, and another particle or hole created during the process exits. 

   We will now discuss the effect of the temporal structure of an approximation to the excitation spectrum which is given by the pole structure of the response function. We know that a time-local kernel only reproduce poles whose locations are given by differences in the renormalized, single-particle excited and ground state energies~\cite{McLachlan1, Casida,Romaniello_JCP130,Sangalli_JCP134}. The excitations corresponding to these poles are then called one-particle excitations, while all remaining ones we refer to as many-particle excitations. However, we can also view these one- to many-particle excitations in terms of some single-particle basis which has an appropriate energy label. In this case, excitations can have one- to many-particle character depending on their representation in this basis. In a wave function based theory, excitations are characterized with help of the basis expansion of the many-particle excited, and ground states. On the other hand, in many-body perturbation theory, we examine the diagrammatic structure of a response function in terms of Green's functions represented in this single-particle basis. 

      \begin{figure}[b]
         \centering
         \subfloat[self-energy insertions]
         {
            \centering
            \includegraphics[scale=1.0]{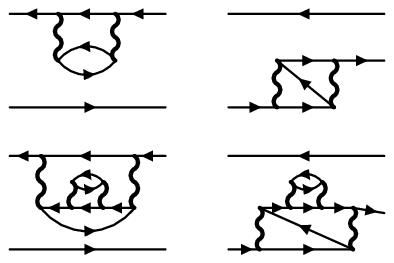}
            \label{fig:excitations_selfenergy}
         }
         \subfloat[kernel contributions]
         {
            \centering
            \includegraphics[scale=1.0]{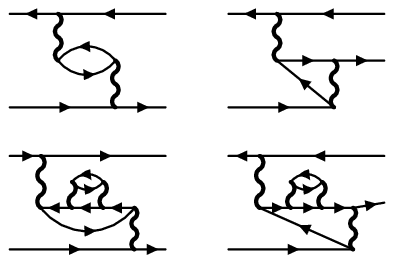}
            \label{fig:excitations_kernel}
         }
         \caption{Some diagrams containing multiple simultaneous particle-hole pairs, whose origin lies either in self-energy insertions or kernel diagrams. A simple line represents a Hartree-Fock Green's function in terms of which the top and bottom rows describe two- and three-particle excitations, respectively. The direction of time is from left to right.}
         \label{fig:many_particle_excitations}
      \end{figure}

      In figure~\ref{fig:many_particle_excitations} we show some low-order diagrams which are obtained by expanding 2nd Born Green's functions in terms of Hartree-Fock Green's functions. These diagrams represent time-nonlocal processes which contain multiple, simultaneous particle-hole pairs, or in other words describe many-particle excitations in the Hartree-Fock basis. Such additional excitations are clearly not present within the Hartree-Fock approximation and can, in general, arise either from self-energy insertions or directly from the kernel as shown in figures~\ref{fig:excitations_selfenergy}~and~\ref{fig:excitations_kernel}, respectively. Moreover, self-consistency ensures that some classes of these diagrams representing many-particle excitations are taken into account to all orders of perturbation theory. If some of these diagrams are dominant in the perturbation expansion, we can expect many-particle excitations to appear in the excitation spectra.


   \section{Model and test systems}
   \label{sec:model}
   We consider $N$-electron lattice systems which are based on the Pariser-Parr-Pople (PPP) model~\cite{Linderberg1} and are described with a Hamiltonian given by
   \begin{eqnarray}
   \label{eq:model_hamiltonian}
      \fl \widehat{H}(t) = \sum_{i\alpha,j\beta} h_{ij}(t)\hat{c}^{\dagger}_{i\alpha}\hat{c}_{j\beta} + U\sum_{i} \hat{n}_{i\uparrow}\hat{n}_{i\downarrow} + \frac{1}{2}\sum_{i \neq j} w_{ij} (\hat{n}_{i}-Z)(\hat{n}_{j}-Z)\, ,
   \end{eqnarray}
   where $\hat{c}_{i\alpha}^{(\dagger)}$ annihilates (creates) an electron of spin $\alpha$ at site $i$, while $\hat{n}_{i} \equiv \hat{c}_{i\sigma}^{\dagger}\hat{c}_{i\sigma}$ and $\hat{n}_{i} \equiv \sum_{\alpha} \hat{n}_{i\alpha}$ are site density operators. The one-body Hamiltonian is given by
   \begin{eqnarray}
      h_{ij}(t) = -\delta_{\langle i,j\rangle} + \delta_{ij}v_{i}(t)
   \end{eqnarray}
   where the hopping elements $\delta_{\langle i,j\rangle}$ are equal to $1$ for the nearest-neighbor sites on the lattice and are zero otherwise. The symbol $v_{i}(t) = v\delta_{1,i}\delta(t)$ denotes a time-dependent potential. The two-body, electron-electron interaction is described with local and nonlocal interaction matrix elements denoted with $U$, known as Hubbard-$U$, and $w_{ij}$, such that $w_{ii} \equiv 0$, respectively. These interaction matrix elements and the positive background potential $Z$ are chosen to address two kinds of systems itemized below.

   \begin{enumerate}
      \item The Hubbard model is specified by choosing the off-diagonal interaction matrix elements, as well as the positive background potential, as zero ($w_{ij} = 0$, $Z = 0$). The model represents a system with a short-range, or more precisely a contact two-body interaction. 
      \item The PPP model, on the other hand, is introduced to address the effects of long-range interactions, which are described by the Coulomb type matrix elements $w_{ij} = U/2d_{ij}$, where is a parameter describing the distance between sites $i$ and $j$. We set the distance between nearest-neighbors on a lattice, as well as the positive background potential to one ($d_{\langle i,j\rangle} = 1, Z=1$). 
   \end{enumerate}

\begin{figure}[t]
      \centering
      \subfloat[spatial profile]
      {
         \label{fig:systems_spatial}
         \centering
         \begin{minipage}[c][120pt][c]{0.4\textwidth}
            \centering
            \includegraphics[scale=0.45]{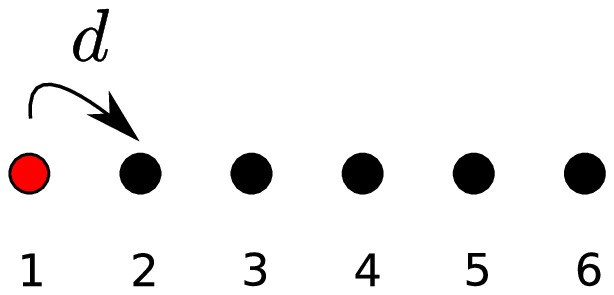}
            \vskip 15pt
            \includegraphics[scale=0.45]{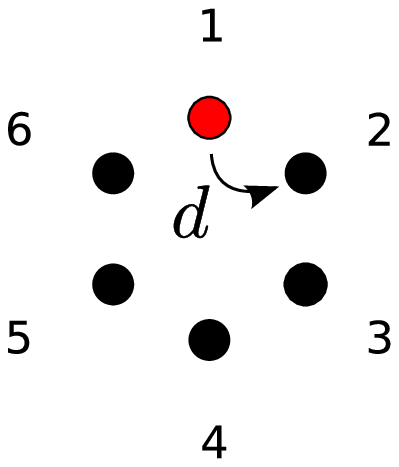}
         \end{minipage}
         }
         \subfloat[Hartree-Fock energies]
         {
         \label{fig:systems_energies}
         \centering
         \begin{minipage}[c][120pt][c]{0.4\textwidth}
            \centering
            \includegraphics[scale=0.4]{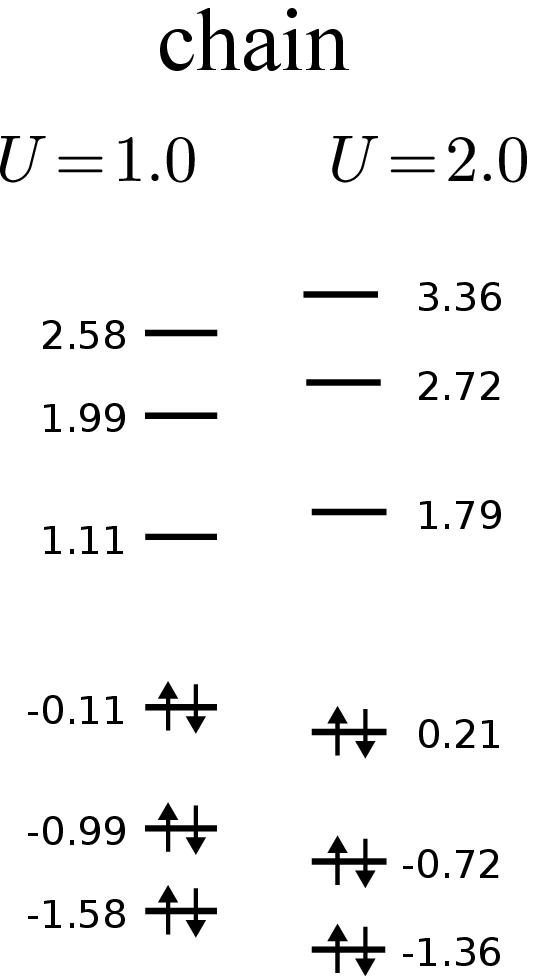}
            \hskip 15pt
            \includegraphics[scale=0.4]{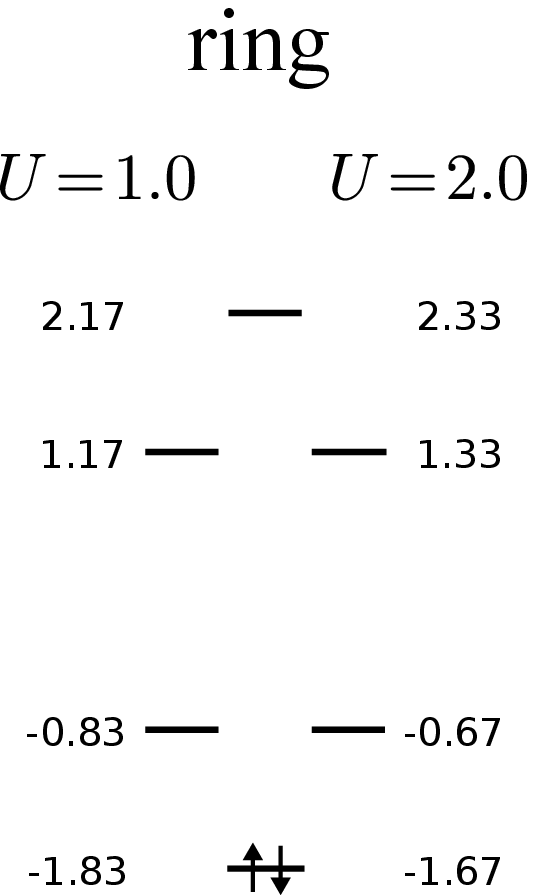}
         \end{minipage}
      }
      \caption{The red sites in the spatial profiles of our test systems, shown in (a), denote the target site for a perturbation, while $d \equiv d_{<ij>} = 1$ is the distance between nearest-neighbors. In (b), we show the Hartree-Fock single-particle energies for two interaction strengths ($U=1.0,2.0$). Notice that the energy spectra of the ring coincide with the non-interacting spectra up to a constant because of the discrete rotational symmetry.}
      \label{fig:systems}
   \end{figure}

   We have, in particular, prepared lattice systems of two main types: linear chains and uniform rings, and conducted calculations by varying the interaction strengths, and number of electrons, but restricted ourselves to spin-compensated systems. In the next section, we summarize some of our results by considering a six site, six electron ($1/2$-filled) chain with long-range interactions, as well as a six site, two electron ($1/6$-filled) ring with short-range interactions. Some essential features of both test systems are represented in figure~\ref{fig:systems}. 

   \section{Results: addition, removal and excitation spectra}
   \label{sec:results}
   We have chosen two finite lattice systems, described in section~\ref{sec:model}, to test the performance of the many-body approximations obtained from the Hartree-Fock and second Born self-energies. The observables used as a measure of the quality of these approximations are the ground state energy, and the non-neutral and neutral excitation spectra calculated using the spectral and retarded response functions, respectively. The quality of our many-body schemes is assessed by comparing all approximate quantities against exact results. 


      \subsection{Numerical aspects}
      All exact quantities are obtained by exact diagonalization (ED), or in other words calculated using the many-particle eigenstates of the system which are obtained by solving the time-(in)dependent Schr\"{o}dinger equation. Our numerical implementation of exact diagonalization in these finite lattice systems relies on the well-known Lanczos routine~\cite{Saad,Lin,Park1}. Approximate quantities, on the other hand, are obtained by calculating the equilibrium Green's function~\cite{Dahlen1,Stan1} which yields the ground state energy, and from which the non-equilibrium Green's function is obtained by by time-propagation. These calculations are performed with a parallelized version~\cite{Balzer1} of the algorithm described in~\cite{Dahlen3,Stan2}.

      Some additional remarks on calculation of spectral properties are in order. These properties are, in practice, computed either by direct evaluation of addition, removal, and excitation energies (only in ED), or by finite-length time-propagation of a many-particle ground state (ED) and an equilibrium Green's function (MBPT). The latter approach leads to time-domain spectral and response functions which are transformed into their frequency-domain counterparts. As a side-effect of the finite propagation length, oscillatory noise-like features, as well as broadening of peaks, appear into these frequency-domain spectra. Such defects can be reduced by increasing the propagation length denoted with $T$ and reached with a time step $\Delta$. In figures, result related to the direct evaluation scheme  will be denoted with ED-D, while time-propagated results are denoted with ED, HF or 2B. 


      \subsection{Addition and removal energies}
      The spectral function relates to so-called quasi-particle properties, in particular, the frequency-domain structure contains information about particle addition and removal energies, and the diagonal elements represent the local density of states. The Lehmann representation of the ground state spectral function is given by 
      \begin{eqnarray}
         \fl A_{i\alpha,j\beta}(\omega) = 2\pi \sum_{n} \bigg(g_{n,i\alpha}^{N+1}{g_{n,j\beta}^{N+1}}^{*} \delta\big(\omega - \Omega_{n}^{N+1}\big) + g_{n,j\beta}^{N-1}{g_{n,i\alpha}^{N-1}}^{*}\delta\big(\omega + \Omega_{n}^{N-1}\big)\bigg)\, ,
      \end{eqnarray}
      where $g_{n,i\alpha}^{N + 1} \equiv \big<\Phi_{0}^{N}\big| \hat{c}_{i\alpha}\big|\Phi_{n}^{N + 1}\big>$ and $g_{n,i\alpha}^{N - 1} \equiv \big<\Phi_{0}^{N}\big| \hat{c}_{i\alpha}^{\dagger}\big|\Phi_{n}^{N - 1}\big>$ are the oscillator strengths, and $\Omega_{n}^{N \pm 1} \equiv E_{n}^{N \pm 1} - E_{0}^{N}$ the addition and removal energies. Symbols $\big|\Phi_{k}^{N}\big>$ and $E_{k}^{N}$ denote the $k$th eigenstate and -energy of a $N$-particle system, respectively. 

\begin{figure}[h]
         \centering
         \includegraphics[width=0.9\textwidth]{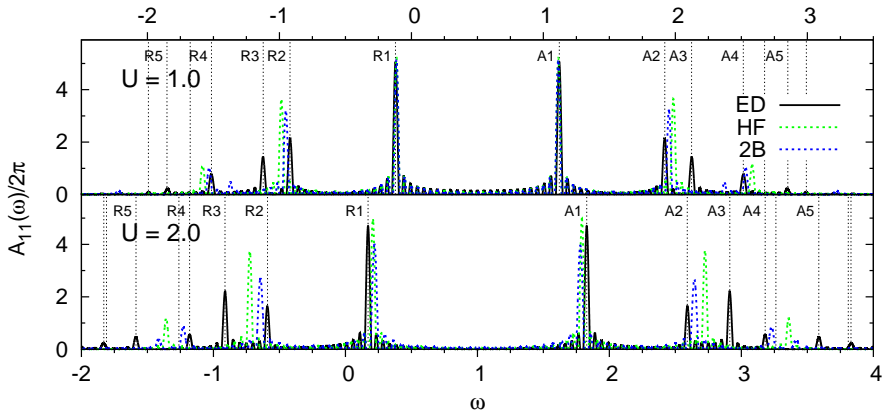}
         \caption{The spectral functions ($A_{11}(\omega) \equiv A_{1\uparrow,1\uparrow}(\omega) \equiv A_{1\downarrow,1\downarrow}(\omega)$) of the six site/electron ($1/2$-filled) chain with long-range (PPP) interactions. Exact addition/removal energies (ED-D) with a non-zero oscillator strength are given by the dashed, vertical lines. ($T = 125, \Delta = 0.02-0.025$)}
         \label{fig:spectral_chain}
      \end{figure}

      The spectral functions of the $1/2$-filled chain with different interaction strengths ($U=1.0,2.0$) are shown in figure~\ref{fig:spectral_chain}. Both many-body approximations capture the lowest addition and removal energies well, irrespective of the interaction strength. Moreover, they open up the band-gap adequately, and reproduce correctly the particle-hole symmetry, which although being in general broken by the long-range interaction, has been restored by the choice $Z=1$ for the positive background. Differences between approximate solutions, as well between approximate and exact spectra, appear at higher energies, see addition energies A2-A5 or removal energies R2-R5, The 2nd Born approximation reproduces some of this additional structure present in the exact spectral function, while the Hartree-Fock method, merely capturing the single-particle energy spectra (Koopmans' theorem), fails to do so. 

\begin{figure}[h]
         \centering
         \includegraphics[width=0.9\textwidth]{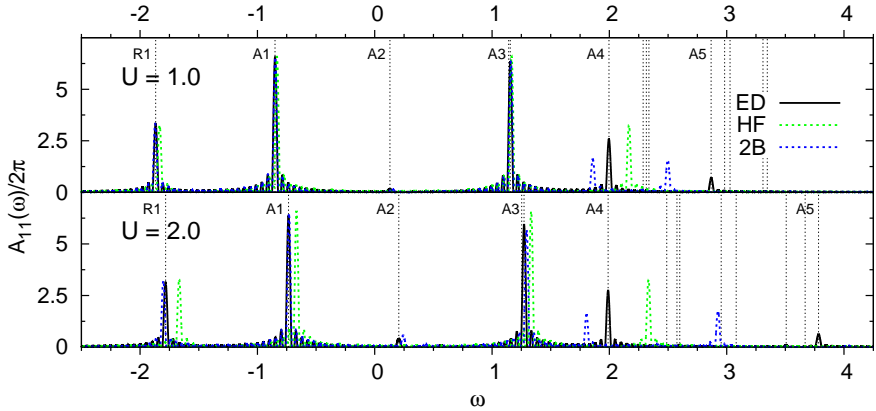}
         \caption{The spectral functions ($A_{11}(\omega) \equiv A_{1\uparrow,1\uparrow}(\omega) \equiv A_{1\downarrow,1\downarrow}(\omega)$) of the six site, $1/6$-filled (two electrons) ring with short-range (Hubbard) interactions. Exact addition/removal energies (ED-D) with a non-zero oscillator strength are given by the dashed, vertical lines. ($T = 125, \Delta = 0.02-0.025$)}
         \label{fig:spectral_ring}
      \end{figure}

      The addition and removal energy spectra of the $1/6$-filled, six site (2 electrons) ring are given in figure~\ref{fig:spectral_ring}. Again, both approximations capture the lowest addition and removal energies well, but now only when $U=1.0$, and reproduce correctly the broken particle-hole symmetry. As the interaction strength is increased, the Hartree-Fock approximation both underestimates the removal energies and overestimates the addition energies which, altogether, signals a failure to reproduce both the 2-electron ground state, and the lowest 3-electron states correctly. The ground state energies given in table~\ref{fig:ground_state_ring} verify these speculations, as well as the fact that both of our approximations describe the 1-particle system correctly, and therefore the removal energies describe solely the quality of the ground state. Although both approximations reproduce the addition energies (A1,A3) with the highest intensities, our correlated approximation gives more accurate results. However, only the 2nd Born approximation is able to quantitatively account for the additional, low-intensity structure which consists of a pole between high-intensity ones (A2) and two higher lying poles (A4,A5).          

      Overall, we have observed that the Hartree-Fock approximation gives quantitatively incorrect high-energy spectra, which is simply due to fact that there are fewer single-particle energies than eigenenergies of the true interacting system. The second Born approximation, on the other hand, capable of accounting for the so-called correlation induced quasi-particles, does reproduce some of warranted additional structure, and additionally captures correctly shifting intensities as interaction strength is increased, see poles A2/R2 of figure~\ref{fig:spectral_chain} or A2 of figure~\ref{fig:spectral_ring}. In the next section, we will address the question of how well these fundamental differences affect the excitation structure evaluated at a higher $\delta \Sigma/\delta G$-level.
 
      \begin{table}[h]
         \centering
         \subfloat[$1/2$-filled, six site, PPP chain]
         {
            \label{fig:ground_state_chain}
            \begin{tabular}{c c c c}
               \br
               $U$ & $E_{0}$ & HF    & 2B     \\ \mr
               1.0 & -6.217  & 0.7\% & 0.1\%  \\ 
               2.0 & −5.540  & 3.2\% & 1.0\%  \\ 
               \br
            \end{tabular}
         }
         \hskip 0.75cm
         \subfloat[$1/6$-filled, six site, Hubbard ring]
         {
            \label{fig:ground_state_ring}
            \begin{tabular}{c c c c}
               \br
               $U$ & $E_{0}$ & HF    & 2B     \\ \mr
               1.0 & -3.867  & 0.9\% & -0.1\% \\
               2.0 & -3.782  & 3.1\% & -0.9\% \\
               \br
            \end{tabular}
         }
         \caption{The exact ground state energies $E_{0}$ and relative errors $E_{HF/2B}/E_{0} -1$, where $E_{HF/2B}$ is the ground state energy in the Hartree-Fock/2nd Born approximation.}
         \label{fig:ground_state}  
      \end{table}


      \subsection{Excitation energy spectra}
      The neutral excitation energies can be accessed through the density(-density) response function, which is given by
      \begin{eqnarray}
      \label{eq:density_response_function}
         \chi_{i\alpha,j\beta}(t,t') \equiv \chi_{i\alpha i\alpha,j\beta j\beta}^{R}(t,t')\, ,
      \end{eqnarray}
      where the retarded response function is defined in equation~\eref{eq:retarded_response_function}. In the following, we consider time-translationally invariant systems in which the response function depends only on a relative time, that is $\chi_{i\alpha,j\beta}(t,t') = \chi_{i\alpha,j\beta}(t-t')$. The connection to the excitation energies becomes explicit in the Lehmann representation, in which the imaginary part of the ground state density response function can be written as 
      \begin{eqnarray}
      \label{eq:density_response_function_lehmann}
         \imag \big(\chi_{i\alpha,j\beta}(\omega)\big)   
         = -\pi\sum_{n} f_{n,i\alpha}f_{n,j\beta} \bigg(\delta\big(\omega - \Omega_{n}\big) - \delta\big(\omega + \Omega_{n}\big)\bigg)\, ,
      \end{eqnarray}
      where $f_{n,i\alpha} \equiv \big<\Phi_{0}^{N}\big| \hat{n}_{i\alpha}\big|\Phi_{n}^{N}\big>$ is the oscillator strength, and $\Omega_{n} \equiv E_{n}^{N} - E_{0}^{N}$ the neutral excitation energies. Such equation is valid as long as the eigenstates of the $N$-particle system, and hence the oscillator strengths, can be chosen real valued, which is the case for our time-reversal invariant systems. 

      However, in practice, we calculate the ground state density response function $\chi_{ij}(t) \equiv \sum_{\alpha\beta} \chi_{i\alpha,j\beta}(t)$ by means of time-propagation. First, we perturb the system with a spatially and temporally local potential $v_{i}(t) = v_{j}\delta_{ij}\delta(t)$, where $j$ denotes the target site for the perturbation, $v$ is the strength of the perturbation and $\delta(t)$ a Dirac delta function. Then, we use equation~\eref{eq:linear_response_relation} which states that the density response function is given by
      \begin{eqnarray}
        \chi_{ij}(t) = \frac{n_{i}(t)-n_{i}^{0}}{v_{j}} + \mathcal{O}(v_{j})\, ,
      \end{eqnarray}
      where $n_{i}^{0}$ denotes the ground state site density, and $n_{i}(t) = \sum_{\alpha}\gamma_{i\alpha,i\alpha}(t)$ the site density of the perturbed system evaluated at time $t$. Although nonlinear effects are always present, they are easily reduced by ensuring that the magnitude of the perturbation lies well in the linear response region. In practice this is verified by doubling the perturbation, and checking that also the response doubles to a sufficient accuracy.

      Before presenting neutral excitation spectra, we will introduce some auxiliary quantities which are useful for the discussion of the results. 
      \begin{enumerate}
         \item
         The ground state energy can be used to judge whether a possible deviation from an exact excitation energy results from a failure to describe the ground, or excited state correctly. Although we cannot access the ground state energy at the $\delta \Sigma / \delta G$-level, a first order approximation denoted with $E_{GM}$ is provided by the Galitski-Migdal functional~\eref{eq:galitskimigdal} evaluated with the 2nd Born self-energy. 
         \item
         We prove in~\ref{sec:proof_of_fsum_rule} that any diagrammatically well-defined, self-consistent many-body approximation satisfies the f-sum rule, that is the density response function satisfies equations~\eref{eq:sum_rule_time} and~\eref{eq:sum_rule_frequency}. In our case, this simply means that 
         \begin{eqnarray}
            \label{eq:fsum_rule}
            c_{ij}^{1} = c_{ij}^{1,t} = c_{ij}^{1,\omega} \, ,
         \end{eqnarray}
         where $c_{ij}^{1,(t/\omega)} \equiv \sum_{\alpha\beta} c_{i\alpha,j\beta}^{1,(t/\omega)}$. On the left hand side of this equation, we have a commutator (see equation~\eref{eq:sum_rule_commutator_exact}) defined as
         \begin{eqnarray}
            \label{eq:fsum_commutator}
            c_{ij}^{1} &\equiv h_{ji}\gamma_{ij} + \gamma_{ji}h_{ij} - \delta_{ij}\sum_{k}\big(h_{jk}\gamma_{kj} + \gamma_{jk}h_{kj}\big)\, ,
         \end{eqnarray}
         where $h_{ij}$ and $\gamma_{ij} \equiv \sum_{\alpha\beta} \gamma_{i\alpha,j\beta}$ are the ground state, one-body Hamiltonian and reduced density matrix, respectively. Whereas, on the right hand side, we have the first time-derivative and frequency-moment of the density response function which are given, respectively, by
         \numparts
            \begin{eqnarray}            
               \label{eq:fsum_derivative}
               c_{ij}^{1,t} &\equiv -\partial_{t} \chi_{ij}(t)\big|_{0^{+}} \, , \\
               \label{eq:fsum_moment}
               c_{ij}^{1,\omega} &\equiv -\frac{1}{\pi}\int\limits_{-\infty}^{\infty}\!d\omega\; \omega \imag \big(\chi_{ij}(\omega)\big) \, .
            \end{eqnarray}
         \endnumparts
         We have also checked numerically, by fixing a method (ED, HF, 2B) and comparing the commutator~\eref{eq:fsum_commutator} against the time-derivative~\eref{eq:fsum_derivative}, that the f-sum rule is satisfied up to a numerical accuracy better than $0.01\%$ for all of our methods (ED, HF, 2B). Moreover, since our response functions are obtained by finite length time-propagations, we have reserved the comparison between the commutator~\eref{eq:fsum_commutator} and frequency moment~\eref{eq:fsum_moment} as a test of the completeness of our excitation spectra. 

         As the frequency moment should equal to the commutator, we can use the latter to estimate how well an approximate method is able to reproduce the true spectra. In table~\ref{fig:frequency_sum}, we compare exact (ED) and approximate (HF, 2B) ground state commutators~\eref{eq:fsum_commutator}. Such comparison shows that the second Born captures, in all cases, more precisely the value of the exact commutator, and is therefore expected to reproduce better quality spectra than the Hartree-Fock.   

         \begin{table}[h]
            \centering
            \subfloat[$1/2$-filled, six site, PPP chain]
            {
               \begin{tabular}{c c c c}
                  \br
                  U   & $c^{1}$ & HF     & 2B     \\ \hline
                  1.0 & 1.755  & 1.1\% & 0.4\%    \\
                     2.0 & 1.733  & 3.9\% & 2.1\% \\
                     \br
                  \end{tabular}
            }
            \hskip 0.75cm
            \subfloat[$1/6$-filled, six site, Hubbard ring]
            {
               \begin{tabular}{c c c c}
                  \br
                  U   & $c^{1}$ & HF     & 2B    \\ \hline
                  1.0 & 1.324  & 0.7\% & -0.2\%  \\
                  2.0 & 1.307  & 2.1\% & 1.2\%   \\
                 \br
               \end{tabular}
            }
            \caption{The relative error $c_{X}^{1}/c^{1} -1$, where $c_{(X)}^{1} \equiv c_{(X);11}^{1}$ and $X=$HF or 2B.}
            \label{fig:frequency_sum}
         \end{table}
         \item
         Lastly, we need some means to characterize excitations which appear in the exact excitation spectra. As the expectation value of an occupation number operator contains information about single-particle state occupations in a compact statistical format, it is a suitable tool for the task. Our occupation number operator is given by 
         \begin{eqnarray}
            \hat{n}_{i}^{\mathrm{HF}} \equiv \sum_{\alpha}\hat{d}_{i\alpha}^{\dagger}\hat{d}_{i\alpha}\, ,
         \end{eqnarray} 
         where $\hat{d}_{i\alpha}^{(\dagger)}$ annihilates (creates) an electron from (to) the single-particle eigenstate \{$i,\alpha$\} of the Hartree-Fock Hamiltonian. We can obtain information about single-particle transitions by comparing the expectation values of this operator for the ground and excited states of the many-particle system. However, since excitation spectra can comprise degenerate excited states which have varying spectral intensities, we need a construct which can distinguish the character of an excitation in such a situation. For this purpose, we define the weighted occupation numbers
         \begin{eqnarray}
            N_{ij,k}^{\mathrm{HF}}(E) \equiv \Tr \big(\rho_{ij}(E) \hat{n}_{k}^{\mathrm{HF}}\big)\, , 
         \end{eqnarray}
         where $\Tr$ is the trace over a complete set of many-particle states, and 
         \begin{eqnarray}
            \hat{\rho}_{ij}(E) 
            \equiv \frac{\sum_{k,E_{k}=E}\sum_{\alpha\beta}f_{k,i\alpha}f_{k,j\beta}\big|\Phi_{k}^{N}\big>\big<\Phi_{k}^{N}\big|}{\sum_{k,E_{k}=E}\sum_{\alpha\beta} f_{k,i\alpha}f_{k,j\beta}}\, ,
         \end{eqnarray}
         a properly normalized density operator which projects any state to a degenerate subspace, with eigenvalue $E$, of the many-body eigenstates. The oscillator strength $f_{k,i\alpha}$ is defined below equation~\eref{eq:density_response_function_lehmann}. In practice, we calculate the differences 
         \begin{eqnarray}
            \Delta N_{ij,k}^{\mathrm{HF}}(E) \equiv N_{ij,k}^{\mathrm{HF}}(E) - N_{ij,k}^{\mathrm{HF}}(E_{0})\, ,
         \end{eqnarray}
         where $E_{0}$ is the true ground state energy. This quantity tells us what kind of single-particle transitions are involved in the many-particle transitions from the ground state to the excited states with energy $E$. For example, consider a spin-compensated two-particle and -level system with nearly one-determinental ground state. Then weighted occupation numbers for the lowest single-particle state whose values are close to one or two describe one- or two-particle excitations, respectively.  
      \end{enumerate}

      \begin{figure}[b]
         \centering
         \includegraphics[width=0.9\textwidth]{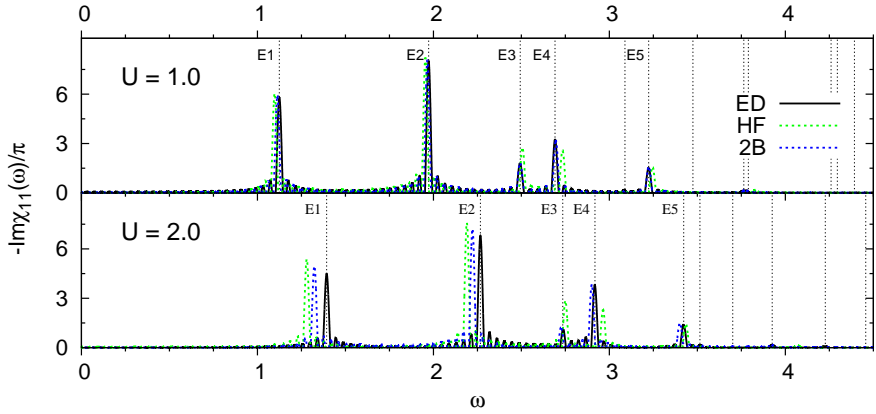}
         \caption{The imaginary part of the density response function $\chi_{11}(\omega)$ of a six site/electron ($1/2$-filled) chain with long-range (PPP) interactions, for which the f-sum rule is satisfied up to $99.9-100.2\%$ when $c_{11}^{1,\omega}$ is used. Exact excitation energies with a non-zero oscillator strength (ED-D) are given by the dashed, vertical lines. ($v_{1} = 0.01,T = 150, \Delta = 0.02-0.025$)}
         \label{fig:density_response_chain}
      \end{figure}

\begin{figure}[h]
         \centering
         \subfloat[Excitations of the $1/2$-filled chain at $U=1.0$, see figure~\ref{fig:density_response_chain}]
         {
            \label{fig:weighted_occupations_chain}
            \includegraphics[height=0.85\textwidth,angle=-90]{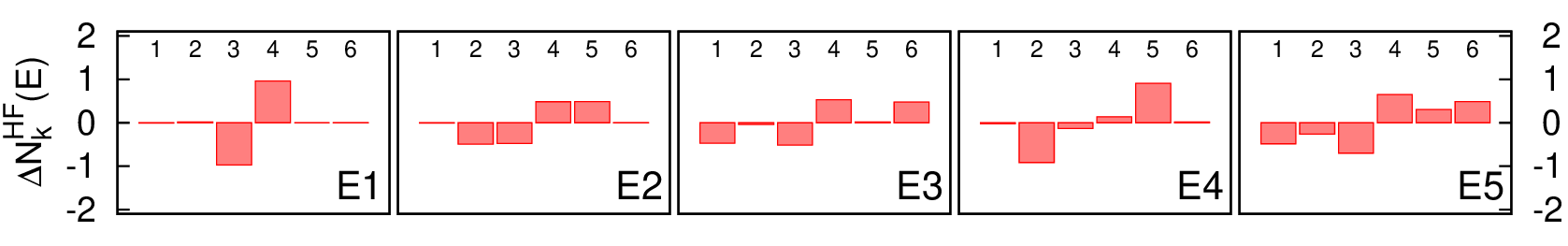}
         }
         \vskip 0.0cm   
         \subfloat[Excitations of the $1/6$-filled ring at $U=1.0$, see figure~\ref{fig:density_response_ring}]
         {
            \label{fig:weighted_occupations_ring}
            \includegraphics[height=0.85\textwidth,angle=-90]{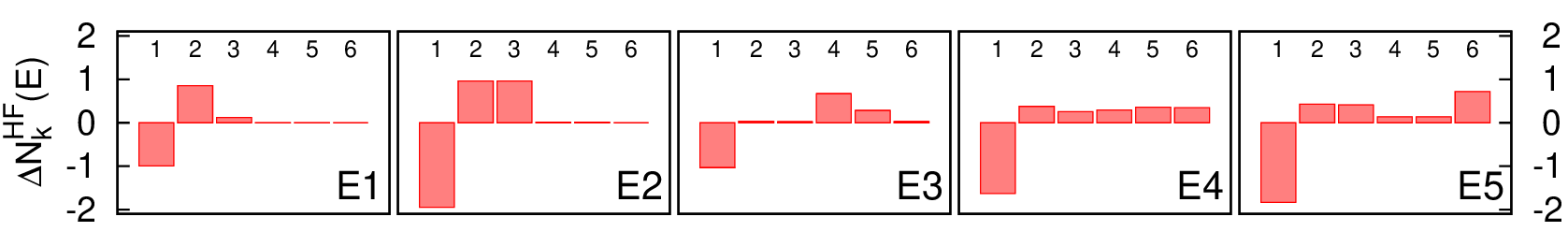}
         }
         \caption{The relative, weighted occupation numbers $\Delta N_{k}(E) \equiv \Delta N_{11,k}^{\mathrm{HF}}(E)$ are shown for each Hartree-Fock state ($k=1,-6$), and excitation E1-E5. The histograms whose filled (positive) or depleted (negative) parts sum up to $0.5-1.5$ or $1.5-2.0$ describe a state with a singly- or doubly-excited character, respectively.}
         \label{fig:weighted_occupations}
      \end{figure}

      The neutral excitation spectrum of the $1/2$-filled chain is shown in figure~\ref{fig:density_response_chain} for two ($U=1.0,2.0$) interaction strengths. The low interaction spectra (see the upper panel) are very similar, that is both many-body approximations reproduce the intensities and frequency structure of the exact response function quite accurately, although the 2nd Born approximation slightly better. As the interaction strength is increased (see the lower panel) both many-body approximations still reproduce quantitatively correct results, although some low-energy structure, see excitations E1 and E2, has started to shift towards the lower-end of the spectrum. This shift is a possible sign of a failure to describe the true ground state, as backed-up by the fact that the squared overlap of the Hartree-Fock ground state with the true ground state is only $0.86$. However, since the correction $E_{GM} - E_{0} \approx 0.06$ is small, such shift cannot be attributed solely as a breakdown of ground state. Overall, dominant features, like the intensities, and energies of low-energy excitations, as well as finer details, such as the diminishing intensity of the third dominant pole (E3) of the response function as a function of the interaction, are replicated more accurately by the correlated approximation, while the mean-field picture even fails to capture the subtleties. The fact that even the Hartree-Fock method reproduces all five, dominant excitation energies (E1-E5) correctly implies that we are dealing with one-particle excitations, as verified by the weighted occupation numbers of figure~\ref{fig:weighted_occupations_chain}. We will shortly consider the question how do the many-body approximations cope with a more complicated excitation spectrum, in which additional excitations of many-particle nature arise. 

\begin{figure}[h]
         \centering
         \includegraphics[width=0.9\textwidth]{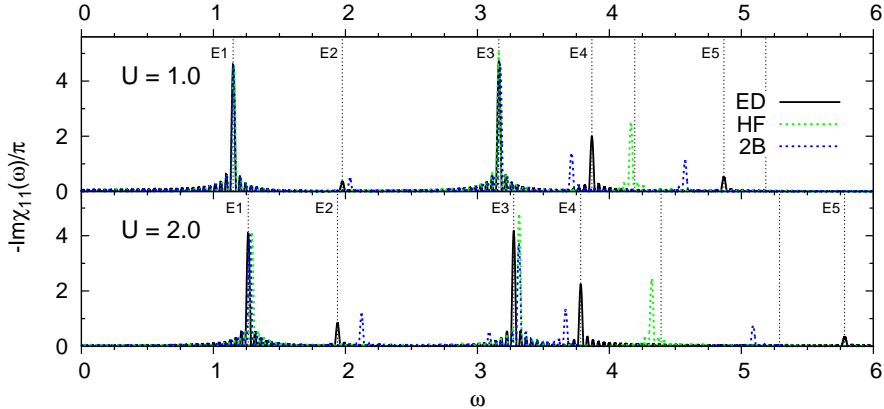}
         \caption{The imaginary part of the density response function $\chi_{11}(\omega)$ of a six site, $1/6$-filled (two electron) ring with short-range (Hubbard) interactions, for which the f-sum rule is satisfied up to $99.8-100.1\%$ when $c_{11}^{1,\omega}$ is used. Exact excitation energies with a non-zero oscillator strength (ED-D) are given by the dashed, vertical lines. ($v_{1}=0.01; T = 150, \Delta = 0.02-0.025$)}
         \label{fig:density_response_ring}
      \end{figure}

      We give an example of such a system by introducing a less rigid, but energetically more simple, $1/6$-filled, six site ring, whose neutral excitation spectrum is given in figure~\ref{fig:density_response_ring} for two ($U=1.0,2.0$) interaction strengths. The results portray, irrespective of the interaction, an adequate agreement between the exact and second Born excitation spectra, but on the contrary, an utter failure of the Hartree-Fock approximation to account correctly for the structure of the response function. Two excitations, one at a lower energy (E2) and another at a higher (E4-E5), which are completely absent in Hartree-Fock, indicate the presence of two-particle excitations, which cannot be described with a time-local approximation. The weighted occupation numbers, shown in figure~\ref{fig:weighted_occupations_ring}, confirm these speculations: only two excitations, E1 and E3, relate to singly-excited states, while all others, in particular E2,E4 and E5, have a strong doubly-excited state character. Therefore, the highest excitation of Hartree-Fock attempts to mimic a practically non-existent lowest to highest state transition, see figure~\ref{fig:systems_energies}, while the 2nd Born approximation successfully reproduces the two-particle excitations, the lowest one being quite accurate, while the higher ones are shifted towards lower energies. As a disadvantage of the correlated scheme, a possibly spurious excitation appears on left-hand side of E3 when $U=2.0$. Such non-physical defects have also been seen in other recent studies~\cite{Romaniello_JCP130, Sangalli_JCP134}, where they are associated with a response kernel which does not have a proper mathematical structure.

   \section{Summary and conclusions}
   \label{sec:conclusions}
   We have studied the performance of the time-dependent Hartree-Fock and second Born approximations by investigating the spectral properties of some finite, strongly correlated, lattice systems. Our purpose has been two-fold: firstly, to test approximations used in transient quantum transport, and secondly to gain insight into excitations in MBPT, both using the Kadanoff-Baym equations. We have calculated ground state energies, as well as spectral and density response functions with our approximate methods, and compared these against exact results, which were obtained by exact diagonalization. 

   The results show that both approximations perform well in a simple half-filled system. That is, they reproduce correctly the true excitation spectra, although the quasi-particle properties are not replicated as accurately, especially in the Hartree-Fock approximation. However, at a lower filling, we observe two-particle excitations, which by construction cannot be captured with the Hartree-Fock method, but are generated correctly in the second Born approximation. Such a difference is also seen in the spectral function in which additional quasi-particles are formed in this correlated approximation. Overall, the second Born approximation is consistently in a better agreement with the exact results, although signs of spurious excitations, also in more asymmetric and strongly interacting systems, call for further research. 

   We conclude that a time-local approximation can be inadequate for highly correlated lattice systems at a low-filling, which should have an impact on quantum transport, as particle number on device varies. On the contrary, already the simplest time-nonlocal, self-consistent approximation performs more reliably. More complicated, but computationally realizable, many-body approximations, as well as investigation of the effects of self-consistency, remain as motives for a future exposition.  
   

   \ack
   We acknowledge CSC – IT Center for Science Ltd. for the allocation of computational resources, and Academy of Finland for financial support.


   \appendix
   \section{Thomas-Reiche-Kuhn/f-sum rule}
   \label{sec:proof_of_fsum_rule}
   The exact density(-density) response function satisfies a consistency relation known in atomic physics as Thomas-Reiche-Kuhn (TRK) sum rule, or in condensed matter physics as the f-sum rule. It is known~\cite{VanLeeuwen} that the TRK/f-sum rule holds in the spin-position basis for self-consistent MBPT. However, there does not exist a proof in the literature that it is satisfied in a lattice system, in which locality gets redefined and the notion of electromagnetic gauge invariance becomes less transparent. In the following, we prove that a response function obtained using equation~\eref{eq:linear_response_relation} fulfills the TRK/f-sum rule also in a lattice system. 

   The electromagnetic gauge freedom is described in real-space by the transformation $(v(\boldsymbol{r},t),\boldsymbol{A}(\boldsymbol{r},t)) \rightarrow (v(\boldsymbol{r},t) - \partial_{t} \Lambda(\boldsymbol{r},t), \boldsymbol{A}(\boldsymbol{r},t) + \nabla \Lambda(\boldsymbol{r},t))$, where $(v(\boldsymbol{r},t),\boldsymbol{A}(\boldsymbol{r},t))$ and $\Lambda(\boldsymbol{r},t)$ are the gauge potential and function, respectively. We can extend this transformation to a lattice system by introducing a new gauge function $\Lambda_{i}(z)$, where $i$ is a collective site and spin index. This function is defined on the Keldysh contour of figure~\ref{fig:keldysh_contour} on which it satisfies the boundary conditions $\Lambda_{i}(t_{0}) = \Lambda_{i}(t_{0}-\rmi\beta)$. Now, we define the gauge transformation $\hat{h}(z) \rightarrow \hat{h}^{\Lambda}(z)$ where
   \begin{eqnarray}
   \label{eq:hamiltonian_gauge_transform}
      \hat{h}^{\Lambda}(z)
      \equiv& \sum_{ij} \rme^{\rmi (\Lambda_{i}(z)-\Lambda_{j}(z))} h_{ij}(z) \hat{c}^{\dagger}_{i}\hat{c}_{j} - \sum_{i}\partial_{z}\Lambda_{i}(z) \hat{n}_{i}
   \end{eqnarray}
   is the gauge transformed one-body part of the Hamiltonian. Moreover, using the equation of motion~\eref{eq:kadanoff_baym}, we can show that
   \begin{eqnarray}
   \label{eq:gauged_greens_function}
      G_{ij}^{\Lambda}(z,z') = \rme^{\rmi \Lambda_{i}(z)} G_{ij}(z,z') \rme^{-\rmi\Lambda_{j}(z')}\, .
   \end{eqnarray}
   Here, we found essential that the interaction matrix elements are two-index quantities, since then the exponential phase factors $\exp(\rmi \Lambda_{i}(z))$ cancel at each interaction vertex. This cancellation guarantees that 
   \begin{eqnarray}
      \Big(\boldsymbol{\Sigma}[\boldsymbol{G}^{\Lambda}]\Big)_{ij}(z,z') = \rme^{\rmi \Lambda_{i}(z)} \Big(\boldsymbol{\Sigma}[\boldsymbol{G}]\Big)_{ij}(z,z') \rme^{-\rmi\Lambda_{j}(z')}\, ,
   \end{eqnarray}
   which in practice is only true if the one-body Green's function is calculated self-consistently. The Hamiltonian~\eref{eq:hamiltonian_gauge_transform} is given, to first order with respect to $\Lambda_{i}(z)$, by 
   \begin{eqnarray}
      \boldsymbol{h}^{\Lambda}(z) = \boldsymbol{h}(z) + \boldsymbol{v}(z) + \cdots 
   \end{eqnarray}
   where we defined the one-body potential
   \begin{eqnarray}
      v_{ij}(z) \equiv  \rmi h_{ij}(z) \big(\Lambda_{i}(z)-\Lambda_{j}(z)\big) - \delta_{ij}\partial_{z}\Lambda_{i}(z) \, .
   \end{eqnarray}
   Consequently, the first order variation of the one-body Green's function can be written as
   \begin{eqnarray}
   \label{eq:variation_of_greens_function}
      \fl \delta G_{ij}^{\Lambda}(z,z') 
      = \sum_{kl}\int\limits_{\gamma}\!d\bar{z}\; L_{ij,lk}(zz',\bar{z}) v_{kl}(\bar{z}) \nonumber\\
      = \rmi \sum_{kl}\int\limits_{\gamma}\!d\bar{z}\; \bigg( L_{ij,kl}(zz',\bar{z})h_{lk}(\bar{z}) - L_{ij,lk}(zz',\bar{z})h_{kl}(\bar{z}) \nonumber\\
      -\rmi\delta_{kl}\partial_{\bar{z}}L_{ij,lk}(zz',\bar{z}) \bigg) \Lambda_{l}(\bar{z})\, ,
   \end{eqnarray}
   where $L_{ij,kl}(zz',\bar{z})$, is the generalized response function defined in equation~\eref{eq:generalized_response_function}. The second line above was obtained by using partial integration, as well as the boundary conditions $L_{ij,kl}(zz',t_{0}-\rmi\beta) = L_{ij,kl}(zz',t_{0})$ and $\Lambda_{i}(t_{0}-\rmi\beta) = \Lambda_{i}(t_{0})$. We can also expand equation~\eref{eq:gauged_greens_function} to first order with respect to $\Lambda_{i}(z)$, and equate the result against the variation~\eref{eq:variation_of_greens_function}. This leads to a result given by
   \begin{eqnarray}
   \label{eq:ward_identity}
      \fl \big(\delta_{ik}\delta(z',z'') - \delta_{jk}\delta(z,z'')\big) G_{ij}(z,z') \nonumber\\ 
      \hspace{-2.0cm} = \sum_{l}\Big(h_{kl}(z'')L_{ij,lk}(zz',z'') - L_{ij,kl}(zz',z'') h_{lk}(z'')\Big) - \rmi\partial_{z''}L_{ij,kk}(zz',z'')\, ,
   \end{eqnarray}
   which is nothing but the generalized Ward identity relating the generalized response function to the one-body Green's function. At the limit $z'=z^{+}$ using equation~\eref{eq:generalized_retarded_response_functions}, the Ward identity can be split into three real-time equations, one for each Keldysh component. These equations are given by
   \numparts
      \begin{eqnarray}
         &\big(\delta_{ik} - \delta_{jk}\big) \gamma_{ij}(t) 
         = \rmi\big(\chi^{>}_{ij,kk}(t,t) - \chi^{<}_{ij,kk}(t,t)\big)\, , \label{eq:ward_identity_a} \\
         &\rmi\partial_{t'}\chi^{\gtrless}_{ij,kk}(t,t')
         = \sum_{l}\Big(h_{kl}(t')\chi^{\gtrless}_{ij,lk}(t,t')  - \chi^{\gtrless}_{ij,kl}(t,t') h_{lk}(t')\Big)\, , \label{eq:ward_identity_b}
      \end{eqnarray} 
   \endnumparts
   where $\gamma_{ij}(t)$ and $\chi_{ij,kl}^{\gtrless}(t,t')$ denote the one-body reduced density matrix and greater/lesser component of the real-time response function which are defined in equation~\eref{eq:density_matrix} and~\eref{eq:components_of_response_function}, respectively. Consequently, the density response function
   \begin{eqnarray}
      \chi_{ij}(t,t') = \theta(t-t')\big(\chi_{ii,jj}^{>}(t,t')-\chi_{ii,jj}^{<}(t,t')\big)
   \end{eqnarray}
   satisfies an equation given by
   \begin{eqnarray}
      \fl \rmi\partial_{t'} \chi_{ij}(t,t') = \delta(t-t')\big(\chi_{ii,jj}^{<}(t,t')-\chi_{ii,jj}^{>}(t,t')\big) + \theta(t-t')\sum_{k}\Big(h_{jk}(t')\big(\chi^{>}_{ii,kj}(t',t) \nonumber\\
      - \chi^{<}_{ii,kj}(t',t)\big) - \big(\chi^{>}_{ii,jk}(t',t) - \chi^{>}_{ii,jk}(t',t)\big)h_{kj}(t')\Big)\, ,
   \end{eqnarray}
   where we used the Ward identity~\eref{eq:ward_identity_b}. Moreover, evaluating this equation at equal times $t' = t - \eta$ with the limit $\eta \rightarrow 0^{+}$ taken afterwards and using the symmetry $\chi_{ij,kl}^{>}(t,t') = \chi_{kl,ij}^{<}(t',t)$ (see equation~\eref{eq:components_of_response_function}) leads to 
   \begin{eqnarray}
      \fl \rmi\partial_{t'} \chi_{ij}(t,t')\big|_{t'=t^{-}} \nonumber\\
      \hspace{-2.0cm}= \sum_{k}\Big(h_{jk}(t)\big(\chi^{>}_{kj,ii}(t,t) - \chi^{<}_{jk,ii}(t,t)\big) - \big(\chi^{>}_{kj,ii}(t,t) - \chi^{>}_{jk,ii}(t,t)\big)h_{kj}(t)\Big)\, .
   \end{eqnarray}
   If the initially unperturbed system is time-translationally invariant then the density response function depends only on a relative time, or in other words $\chi_{ij}(t,t') = \chi_{ij}(t-t')$. Then the Ward identity~\eref{eq:ward_identity_a} allows us to write the Thomas-Reiche-Kuhn sum, or the f-sum, rule in time-domain as
   \begin{eqnarray}
   \label{eq:sum_rule_time}
      -\partial_{t} \chi_{ij}(t)\big|_{t=0^{+}} 
      =& h_{ji}\gamma_{ij} + \gamma_{ji}h_{ij} -\delta_{ij}\sum_{k}\big(h_{jk}\gamma_{kj} + \gamma_{jk}h_{kj}\big)\, .
   \end{eqnarray}
   Furthermore, by assuming that the density response function is analytic in the upper half of the complex plane, the TRK/f-sum rule can be written in the frequency-domain~\cite{VanLeeuwen} as
   \begin{eqnarray}
   \label{eq:sum_rule_frequency}
      \fl -\frac{1}{\pi}\int\limits_{-\infty}^{\infty}\!d\omega\; \omega \imag\big( \chi_{ij}(\omega)\big) 
      =& h_{ji}\gamma_{ij} + \gamma_{ji}h_{ij} -\delta_{ij}\sum_{k}\big(h_{jk}\gamma_{kj} + \gamma_{jk}h_{kj}\big)\, .
   \end{eqnarray}
   Note that in the text, we refer to the right hand side of these equations as the ground state commutator $c^{1}_{ij}$ since in an exact theory it can be written~\cite{Goodman} as
   \begin{eqnarray}
   \label{eq:sum_rule_commutator_exact}
      c^{1}_{ij} = \big<\Phi_{0}^{N}\big|\Big[\big[\hat{n}_{i},\widehat{H}\big],\hat{n}_{j}\Big]\big|\Phi_{0}^{N}\big> \, ,
   \end{eqnarray} 
   where $\big|\Phi_{0}^{N}\big>$ is the $N$-particle ground state, $\hat{n}_{i}$ the site density operator, and $\widehat{H}(t)$ the many-particle Hamiltonian.


\end{document}